\documentclass[aps,prd,10pt,twocolumn,showpacs,preprintnumbers,amsmath,amssymb,nofootinbib,floatfix]{revtex4-1}

\usepackage{graphicx, bm} 

\usepackage{natbib} %Default!

\usepackage{epsfig,amsmath} 
\usepackage{hyperref} %Default

\usepackage{array}
\usepackage{afterpage}
\usepackage{color}
\usepackage{multirow}

\usepackage{textgreek} %% added to enable upright greek letters for units -SF
\usepackage{etoolbox} %% added to enable toggle if/then switching for \Wcm{18}, 10^18 Wcm2 command -SF

\def\twod/{2D(3$v$)}
\def\um/{\textmu m} % Makes writing micron units easier. E.g., invoke via 55~\um/
\newcommand{\Wcm}[2][0]{\ifnumcomp{0}{=}{#1}{}{$#1~\cdot$~}$10^{#2}$~W~cm$^{-2}$} % Makes writing intensity easier. Invoke via Wcm{18} to write "10^18 W cm-2" or invoke via Wcm[1.2]{12} to write "1.2 x 10^12 W cm-2"
\def\mildI/{\Wcm[5.4]{17}}
\def\relI/{\Wcm[3]{18}}

\begin{document}

%%%%%%% BEGIN UNCOMMENT FOR ARXIV %%%%%%%%%%

\title{3D PIC simulations of electron beams created via reflection of intense \\ laser light from a water target}

\author{Gregory K. Ngirmang$^{1,2,*}$, Chris Orban$^{1,2}$, Scott Feister$^{1,2}$, John T. Morrison$^{2,3}$, Kyle Frische$^{2}$, Enam A. Chowdhury$^{1,4}$, and W.M. Roquemore$^{5}$}

\affiliation{
\vspace{0.2cm}
(1) Department of Physics, The Ohio State University, Columbus, OH \\
(2) Innovative Scientific Solutions, Inc., Dayton, OH \\
(3) Fellow, National Research Council \\
(4) Intense Energy Solutions, Inc., Dayton, OH \\
(5) Air Force Research Laboratory, Dayton, OH 
}

\email{ngirmang.1@osu.edu}
%%%%%%%%%%% END UNCOMMENT FOR ARXIV %%%%%%%%%%%%

% %%%%%%% BEGIN COMMENT OUT FOR ARXIV %%%%%%%%%%%
% \preprint{AIP/123-QED}

% \title{3D PIC simulations of electron beams created via reflection of intense laser light from a water target}

% \author{Gregory K. Ngirmang}
%  \email{ngirmang.1@osu.edu}
%  \affiliation{\mbox{Department of Physics, The Ohio State University, Columbus, OH, 43210, USA}} % mbox used to prevent moving to a separate line
%  \affiliation{Innovative Scientific Solutions, Inc., Plain City, OH, 45459, USA}

% \author{Chris Orban}
%  \affiliation{\mbox{Department of Physics, The Ohio State University, Columbus, OH, 43210, USA}} % mbox used to prevent moving to a separate line
%  \affiliation{Innovative Scientific Solutions, Inc., Plain City, OH, 45459, USA}

% \author{Scott Feister}
%  \affiliation{\mbox{Department of Physics, The Ohio State University, Columbus, OH, 43210, USA}} % mbox used to prevent moving to a separate line
% \affiliation{Innovative Scientific Solutions, Inc., Plain City, OH, 45459, USA}

% \author{John T. Morrison}
% \affiliation{Fellow, National Research Council, Washington, DC, 20001, USA}

% \author{Kyle D. Frische}
% \affiliation{Innovative Scientific Solutions, Inc., Plain City, OH, 45459, USA}

% \author{Enam A. Chowdhury}
% \affiliation{\mbox{Department of Physics, The Ohio State University, Columbus, OH, 43210, USA}} 
% \affiliation{Intense Energy Solutions, LLC., Plain City, OH, 43064, USA}

% \author{W.M. Roquemore}
% \affiliation{Air Force Research Laboratory, WPAFB, OH, 45433, USA}

% %%%%%%% END COMMENT OUT FOR ARXIV %%%%%%%%%%%

\date{\today}

\begin{abstract}
We present 3D Particle-in-Cell (PIC) modeling of an ultra-intense laser experiment by the Extreme Light group at the Air Force Research Laboratory (AFRL) using the PIC code LSP. This is the first time PIC simulations have been performed in 3D for this experiment which involves an ultra-intense, short-pulse (30 fs) laser interacting with a water jet target at normal incidence. These 3D PIC simulation results are compared to results from \twod/ PIC simulations for both \mildI/ and \relI/ intensities. 
Comparing the \twod/ and 3D simulation results, the laser-energy-to-ejected-electron-energy conversion efficiencies were comparable, but the angular distribution of ejected electrons show interesting differences with qualitative differences at higher intensity. An analytic plane-wave model is discussed which provides some explanation for the angular distribution and energies of ejected electrons in the \twod/ simulations. We also performed a 3D simulation with circularly polarized light and found a significantly higher conversion efficiency and peak electron energy, which is promising for future experiments.
%We present 3D Particle-in-Cell (PIC) modeling of an ultra-intense laser experiment by the Extreme Light group at the Air Force Research Laboratory (AFRL) using the PIC code LSP. This is the first time PIC simulations have been performed in 3D for this experiment which involves an ultra-intense, short-pulse (30 fs) laser interacting with a water jet target at normal incidence. These 3D PIC simulation results are compared to results from 2D(3$v$) PIC simulations for both $5.4\cdot10^{17} W cm^{-2}$ and $3\cdot10^{18} W cm^{-2}$ intensities. 
%Comparing the 2D(3$v$) and 3D simulation results, the laser-energy-to-ejected-electron-energy conversion efficiencies were comparable, but the angular distribution of ejected electrons show interesting differences with qualitative differences at higher intensity. An analytic plane-wave model is discussed which provides some explanation for the angular distribution and energies of ejected electrons in the 2D(3$v$) simulations. We also performed a 3D simulation with circularly polarized light and found a significantly higher conversion efficiency and peak electron energy, which is promising for future experiments.
\end{abstract}

\maketitle

%%%%%%%%%%%%%%%%%%%%%%%%%%%%%%%%%%%%%%%%%%%%%%%%%%%%%%%%%%%%%%%%%%%%%%%%%%%%%%%%%%%%%%%%%%

\section{Introduction}
\label{sec:intro}

Ultra-intense laser systems offer a diverse range of potential applications and interrogate a variety of interesting physical regimes. While many research groups focus on highly-relativistic laser intensities where the associated quiver velocity is large ($a_0 \gg 1$), there are still interesting experiments to perform involving laser intensities near \Wcm{18} where the electron dynamics are only moderately relativistic ($a_0 \sim 1$). Importantly, at these intensities, current technology allows few-mJ laser pulses to be created at a kHz repetition rate \cite[e.g.][]{Mordovanakis_etal2009,Morrison_etal2015}, which is advantageous for applications and for accumulating statistics. These intensities can also exhibit high reflectivity ($\gtrsim 70\%$) from near-solid density targets \cite{Panasenko_etal2010}, which is very unlike laser interactions at much higher intensity \cite{Levy_etal2013}. 

%Please Note that Scott objects to us citing Morrison et al. 2015 and saying or implying that the reflectivity was high. Orban et al. 2015 does present reflectivity measurements from simulations.

As discussed in \citet{Orban_etal2015} high reflectivity allows for efficient acceleration of electrons to $\sim$MeV energies. The superposition of the forward and reflected laser pulses create a standing wave pattern of strong electric and magnetic fields can launch electrons in the forward and backwards directions with a relativistic momentum similar to the $a_0$ value of the laser field. This was first noticed by Kemp et al. \cite{Kemp_etal2009} who recognized that this can be used as a mechanism for propelling electrons \emph{forward} into the compressed fuel of an inertial confinement fusion experiment. Orban et al. \cite{Orban_etal2015} studied this phenomenon for laser intensities with $a_0 \sim 1$ for the first time and focused on the \emph{backwards}-directed electrons that can receive an additional energy boost when they are overtaken by the reflected laser pulse. Both the theoretical study of \citet{Orban_etal2015} and the corresponding experimental investigation of \citet{Morrison_etal2015} point to large numbers of MeV electrons being produced from individual few-mJ laser pulses normally incident on a water jet target. Recently, MeV energies were confirmed in a using an electron spectrometer \cite{Feister_etal2015}.

This paper presents the first 3D Particle-in-cell (PIC) simulations of laser-matter interactions for this experiment. In earlier theoretical studies, \citet{Orban_etal2015} relied on \twod/ PIC simulations using the LSP code \cite{Welch_etal2004}. As is well known, \twod/ Cartesian simulations assume symmetry along the vertical dimension, so while there can be currents in all three dimensions (hence the 3$v$) particles cannot move in one of these dimensions as they would in a fully 3D real world. A simple application of Gauss' law to this \twod/ geometry indicates that electrostatic fields decay with a different radial dependence than they would normally. So although \twod/ simulations should be qualitatively accurate, ultra-intense laser-matter interactions are an intrinsically 3D phenomena and must be treated as such.
%Initial Conditions
\begin{figure*}[ht!]
  \includegraphics[angle=0,width=7in]{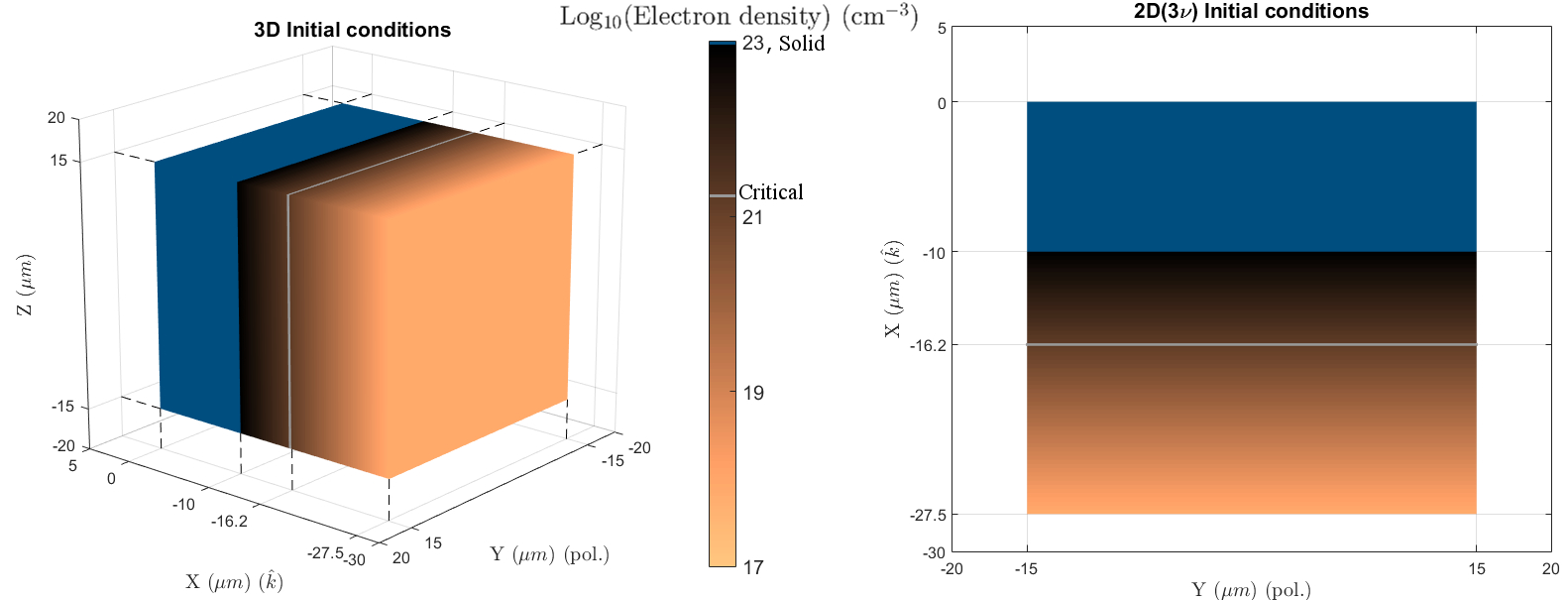}
  \vspace{-0.3cm}
  \caption{Initial conditions of the 3D simulations in the left pane and of the \twod/ simulations in the right pane, specifically the electron density. The target consists of solid density, colored dark blue, that is 10~\um/ deep in the laser axis direction ($x$) and is 30~\um/ along the transverse dimensions. The critical density, which is 1.74$\times$10$^{21}$ cm$^{-3}$, is highlighted in gray on both plots. An exponential density of electrons is placed on the surface facing the laser with a scale length of 1.5~\um/. The laser comes to focus at $(x,y,z)=(0,0,0)$, which is in the solid density of the target for all simulations.}
  \label{fig:IC}
\end{figure*}

While many groups have successfully performed 3D PIC simulations of low density targets (e.g. laser wakefield acceleration), it is especially computationally intensive to perform 3D PIC simulations of near-solid density plasmas as exist in the experiment considered here. There can be energy conservation problems if the Debye length, $\lambda_D$, of the simulated plasma is smaller than the grid resolution \cite{BirdsallLangdon2004} and for near solid density plasmas the Debye length is very small, of order nanometers in scale \cite{NRL}. As discussed in the next section, the LSP code is designed with an implicit algorithm that avoids this grid heating problem. Consequently, one can perform 3D PIC simulations without resolving the Debye length in every area of the grid and still maintain good energy conservation. We present the results of such simulations and compare to earlier \twod/ results throughout this paper. We also address, using analytic methods, aspects of standing wave acceleration with a considerably more sophisticated model than was used in \citet{Orban_etal2015}. 

The initial conditions and configuration for these simulation will be discussed in Sec.~\ref{sec:sims}. Results will be summarized in Sec.~\ref{sec:results} including a comparison to experimental measurements from \citet{Morrison_etal2015}. Finally, a discussion of these results will be presented in Sec.~\ref{sec:discuss}. The Appendix describes an analytic model for electron acceleration in these simulations.

%%%%%%%%%%%%%%%%%%%%%%%%%%%%%%%%%%%%%%%%%%%%%%%%%%%%%%%%%%%%%%%%%%%%%%%%%%%%%%%%%%%%%%%%%%
\section{Particle-In-Cell Simulations}
\label{sec:sims}

We performed 3D PIC simulations with the LSP code\cite{Welch_etal2004}. The initial conditions for these simulations were three dimensional analogs to the simulations described in \citet{Orban_etal2015}. For this paper, we use the following Cartesian coordinate system for these simulations: the positive $x$-axis is the direction of the laser, the $y$-axis is the polarization direction, and $z$-axis is the axis of the water column. As shown in Fig.~\ref{fig:IC}, the target is a 27.5~\um/~$\times$~30~\um/~$\times$~30~\um/
(27.5~\um/ along the laser axis) rectangular block of water-like plasma, consisting of free  electrons, protons, and O$^+$ ions, in proportion to make the target match water's chemical composition and to ensure charge neutrality (O$^+$ to p$^+$ to  e$^-$ ratio of 1:2:3). For 10~\um/ along $x$, the target consists of plasma at solid density with electron density of 1.0$\times$10$^{23}$ cm$^{-3}$, and others in proportion. Beyond 10~\um/ in $x$, we included a decaying pre-plasma profile with an exponential scale length of 1.5~\um/, again along $x$, up to the target's front edge. The target along the The initial density of the target does not vary in the transverse directions. Here, the critical electron density $n_c = 4\pi m \omega^2/e^2$ is 1.74$\times$10$^{21}$ cm$^{-3}$, where $m$ is the mass of an electron, $e$ is elementary charge, $\omega=2\pi\lambda/c$ is angular frequency of the laser, and $c$ is the speed of light. Thus, the solid density target is 57$n_c$. The electron species had a starting temperature of 1 eV.

In these simulations, an 800~nm wavelength laser pulse is normally incident onto the target, well off-focus with the peak focus being at $x = y = z = 0$ in Fig.~\ref{fig:IC} which is as in \citet{Orban_etal2015}. The pulse has a sine-squared envelope with a period of 60 fs, or a FWHM of 30 fs. Amongst the simulations executed, two laser intensities were used, \mildI/ and \relI/. Also, these studies used a linearly polarized laser for both intensities, with an addition 3D simulation with a \mildI/, circularly polarized pulse. We note that the \relI/ and \mildI/ beams had corresponding $a_0$ values at peak focus of 0.5 and 1.2, respectively, where $a_0=e E_0  / \omega m c$, where $E_0$ is the peak electric field of the laser. The incident laser profiles had 2.26~\um/ (2.6~\um/ FWHM) and 2.174~\um/ (2.5~\um/ FWHM) spot sizes for the \mildI/ and \relI/ simulations, respectively. Time steps of 0.1~fs, or about $1/30$th of the laser oscillation period were used for all simulations. Along each dimension, the spatial resolution was 8 cells per wavelength ($\lambda/8\times\lambda/8\times\lambda/8$) for the 3D simulations and 32 cells per wavelength ($\lambda/32\times\lambda/32$) for the correspondent \twod/ simulations. While these simulations do not resolve the Debye length in every cell (since there are cells in the simulation with near-solid densities and nanometer-scale Debye lengths), the phenomena of interest is electron acceleration in the underdense pre-plasma extending from the target where the Debye length is much larger and more easily resolved by the simulation. The implicit algorithm in LSP is designed to avoid grid-heating issues so that the near-solid density regions do not ruin the overall energy conservation of the simulation.  All simulations had 27 macro-particles per cell of each species (free electrons, protons, and O$^+$ ions). As mentioned in \citet{Orban_etal2015}, O$^+$ ions in this simulation are capable of being further ionized by an ionization model in LSP which follows the Ammisov-Delone-Krainov rate \cite{ADK}. Moreover, electrons scatter by a Monte-Carlo algorithm as in \citet{Kemp_etal2004} with a scattering rate determined by a Spitzer model \cite{Spitzer1963}.

The 3D simulations were run for 185~fs or longer; specifically, the \mildI/ and \relI/ simulations were run for 185 fs and 209~fs, respectively, and the corresponding \twod/ sims were run for 250~fs each. The 3D simulations were not evolved to 250~fs to reduce the computational expense and because the total number of $>$120~keV electron macroparticles ejected from the target had already reached a plateau by the time the 3D simulation was ended. An additional 3D simulation was performed with circular polarization but with other parameters being the same as the \mildI/ simulation with linear polarization.

%%%%%%%%%%%%%%%%%%%%%%%%%%%%%%%%%%%%%%%%%%%%%%%%%%%%%%%%%%%%%%%%%%%%%%%%%%%%%%%%%%%%%%%%%%
\section{Results}
\label{sec:results}

% Main Figure
\begin{figure*}[t]
  \includegraphics[angle=0,width=7in]{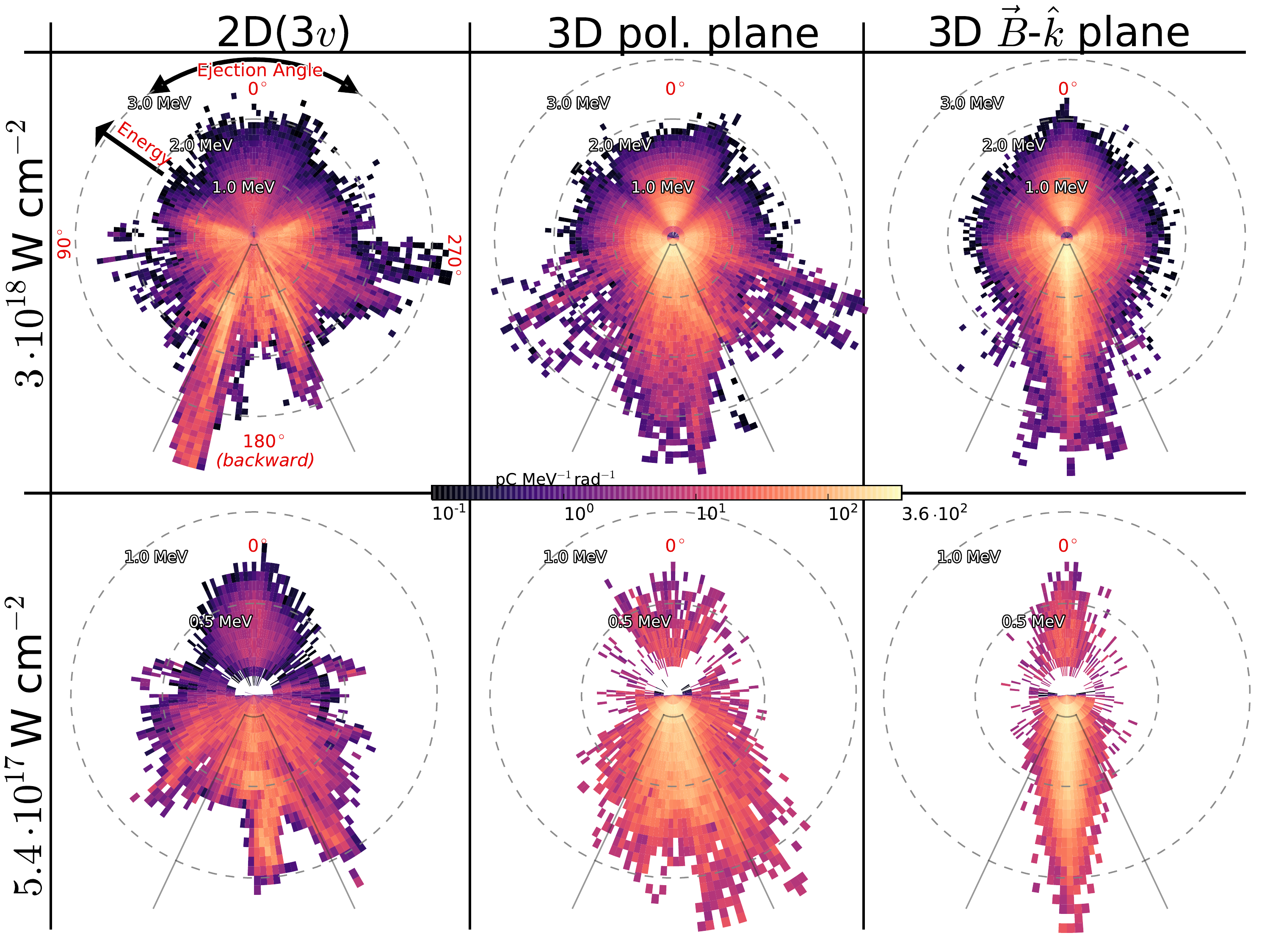}
  \vspace{-0.4cm}
  \caption{
Ejected electron distributions are presented for both \twod/ (left column) and 3D (middle and right column) simulations for \relI/ (upper panels) and \mildI/ (lower panels) intensities with linear polarization. For each plot, distance from the origin indicates the electron kinetic energy, angle from the origin indicates the angle of ejection from the target and color represents the charge per angular-energy bin in pC MeV$^{-1}$ rad$^{-1}$ units. The 3D simulation results show ejected electrons projected both onto the polarization plane (center column) and the perpendicular ($\vec{B}$-$\hat{k}$) plane (right column). A gray line on each plot shows $\pm 25^{\circ}$ of the laser back-reflection axis. Ejected electrons that leave within the conical section defined by these angles are counted in the conversion efficiency measurement (see Table~\ref{table:results}). Electrons with kinetic energies below 120~keV are not counted in this measurement (as discussed in the text) which is why these lines stop at 120~keV in all plots. To directly compare \twod/ and 3D simulations, the charges in the \twod/ simulation (left column) were multiplied by the spot-size of the laser.
%  Ejected electron distributions for two intensities are presented for both \twod/ and 3D simulations with linear polarization. For each plot, distance from the origin indicates electron kinetic energy, angle is the angle of ejection, and color represents the charge density (pC MeV$^{-1}$ rad$^{-1}$).  Angle of ejection is calculated from the final momentum leaving the simulation space, with 180$^\circ$ indicating ejection in the laser back-reflection direction. The 3D simulation results are projected both into the polarization plane and into the perpendicular $\vec{B}$-$\hat{k}$ plane.
%  A gray line on each plot traces a region of $>$120~keV ejection within 25$^{\circ}$ of the laser back-reflection axis. Electrons within this conical section enter the calculation of the laser to electron conversion efficiency in Table \ref{table:results}, as explained in Sec.~\ref{sec:conv}. Note that the radial axis energy range is reduced for the lower intensity plots (bottom).
%  To obtain a correspondence between the 3D and \twod/ simulations, the charges in the \twod/ simulation (1st column) were multiplied by the spot-size of the laser in the corresponding simulation, that is, 2.26 \um/ and 2.174 \um/ for \relI/ (top) and \mildI/ (bottom), respectively.
}
  \label{fig:main_result}
\end{figure*}

\begin{figure*}
  \includegraphics[angle=0,width=6in]{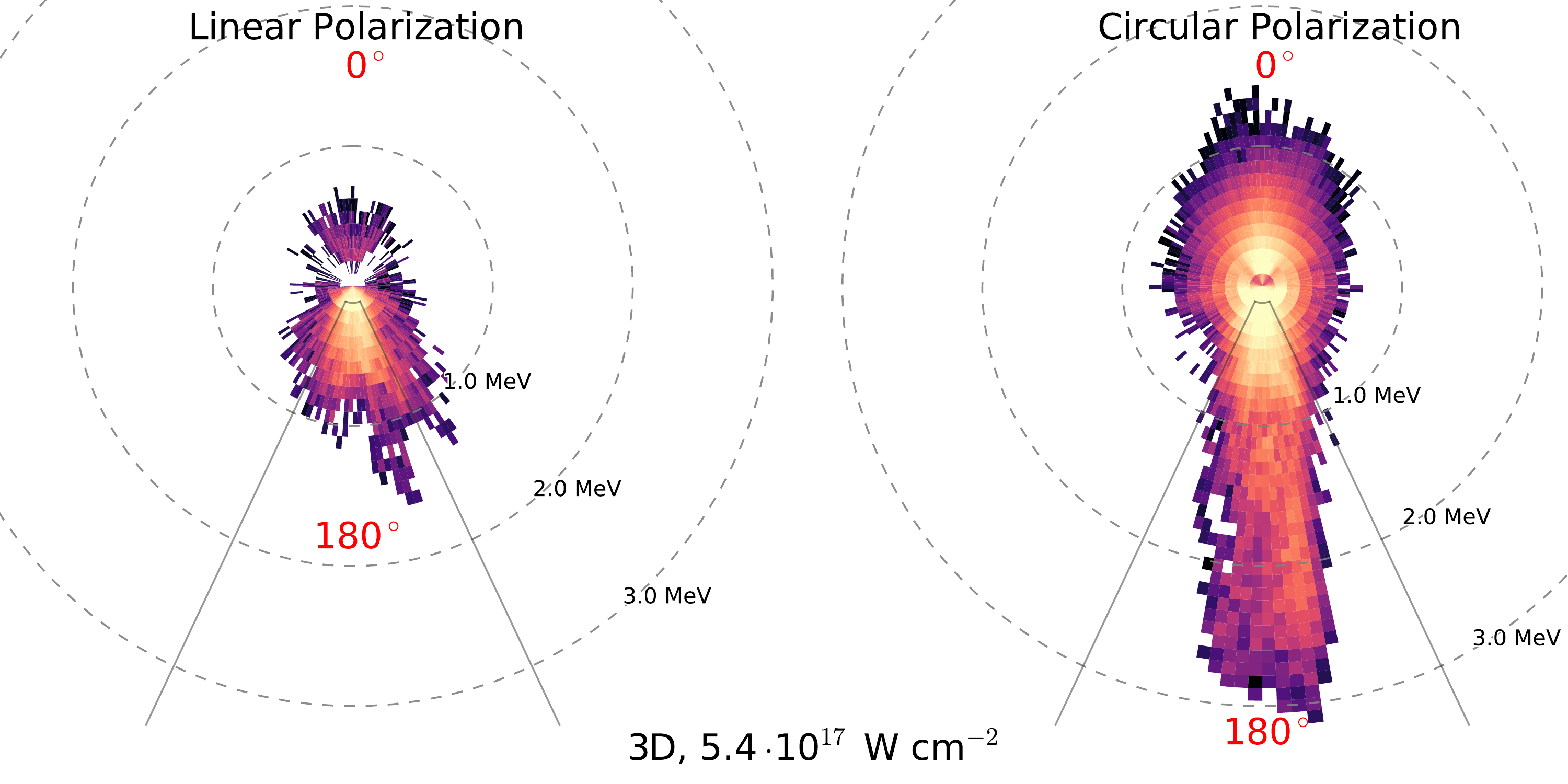}
  \caption{Ejected electron distributions for linearly polarized and circularly polarized lasers with intensity \mildI/. Both plots are results of 3D simulations and show projections onto the $xy$ plane, which in the linearly polarized case is the polarization plane. The plot layout and color mapping are the same as for Fig.~\ref{fig:main_result}.}
  \label{fig:5e17_lpvcp}
\end{figure*}
\subsection{Ejected Charge Ejection-Angle and Energy Spectrum}
\label{sec:e_ang_scpec}

See Fig.~\ref{fig:main_result} for an analysis of the angle and energy distributions of ejected electrons from the linear polarization simulations run. The bins in these plots have a radial (energy) spacing of $\Delta E=50$~keV for the \mildI/ simulations and $\Delta E=100$~keV for the \relI/ simulations. For all plots, angular bins of $\Delta \phi=2^{\circ}$ were used. The ejection angle for each macroparticle was determined from the momentum of the macroparticle as it left the simulation volume. To compare results between the various simulations which warranted different energy spacings, the charge histogram weights are divided by the spacing per cell (reflected in the unit labels above the colorbar). Given that the macro-particles in the \twod/ simulations are in actuality line charges, the ejected macro-particles in \twod/ simulations cannot be compared directly to ejected macroparticles from 3D simulations. Thus, to obtain units of charge instead of linear charge density for the \twod/ plots in the first column of Fig.~\ref{fig:main_result}, every ejected line charge was multiplied by the spotsize of the respective simulation, 2.26 \um/ and 2.174 \um/ for the \relI/ (top) and \mildI/ (bottom), respectively.

Finally, a 3D simulation with the same laser parameters as the \mildI/ simulations was performed with circular (instead of linear) polarization. The ejected electron results for this simulation is shown in Fig.~\ref{fig:5e17_lpvcp} and compared to the results from the simulation with linear polarization.

%defines a column type, using the array package. see
%http://tex.stackexchange.com/questions/5017/center-column-with-specifying-width-in-table-tabular-enviroment
\newcolumntype{x}[1]{>{\centering\arraybackslash}m{#1}}
\begin{table*}
    \begin{tabular}{| x{1.5in} | x{.75in} | x{0.5in} | x{0.5in} | x{0.5in} | x{0.5in} | x{0.5in} |}
    \hline
    \textbf{Intensity} & \Wcm{18} & \multicolumn{3}{c|}{\mildI/} & \multicolumn{2}{c|}{\relI/} \\ \hline
%  \textbf{Simulation/Experiment} &
             &  
  Experiment (Linear Pol.)& 
  2D Linear Pol. & 
  3D Linear Pol. & 
  3D Circular Pol. & 
  2D Linear Pol. &
  3D Linear Pol. \\ \hline
\textbf{Run Time} &
  -- &
  \multicolumn{3}{c|}{209~fs} &
  \multicolumn{2}{c|}{185~fs} \\ \hline
\textbf{Charge Accelerated Backwards ($>$120~keV)} &
  $\sim$ 300~pC &
  -- &
  20.6~pC &
  39.2~pC &
  -- &
  88.5~pC \\ \hline
\textbf{Conversion Efficiency ($>120$~keV)} & 
  $>$ 1.5\% &
  0.71\% &
  0.56\% &
  1.82\% &
  1.42\% &
  1.01\% \\ \hline
  \end{tabular}
    \caption{Summary of the simulation results presented here and experimental results from \citet{Morrison_etal2015}. Conversion efficiencies can be calculated for \twod/ PIC simulations but because of the geometry of these simulations an exact value for the charge ejected cannot be determined as discussed in the text. ``Charge Accelerated Backwards'' is defined as summed charge of electrons ejected from the target with energy $>$120~keV and within an angle of 25$^{\circ}$ from the incoming laser axis. ``Laser Conversion Efficiency'' is defined as the total kinetic energy of these electrons divided by the energy of the incoming laser.}
\label{table:results}
\end{table*}

\subsection{Conversion Efficiency and Total Charge}
\label{sec:conv}

As discussed in \cite{Morrison_etal2015}, the final focusing mirror in the experiment, the so-called Off-Axis Parabola (OAP), was used as a Faraday cup to measure the number of electrons ejected from the target within the solid angle subtended by the mirror. In one experimental configuration a 100~\um/ thick transparent glass slide was placed between the target and the OAP, allowing the laser pulse to pass through but blocking many of the lower energy electrons from reaching the OAP. As discussed in \cite{Morrison_etal2015}, the energy threshold for electrons to pass through the slide is estimated to be about 120~keV. Using this value as a minimum energy for the electrons that are still detected at the OAP gives a robust lower limit on the conversion efficiency from laser energy to electrons with kinetic energies above 120~keV. This lower bound conversion efficiency from experiment can be compared to measurements from simulations. In our simulations the conversion efficiency is inferred in an analogous way to the experiment, counting only the electrons with kinetic energies above 120~keV that are ejected within the angle subtended by the 50.5$^\circ$ degree opening angle (0.60~sr solid angle) of the OAP. Experimental and simulated conversion efficiencies are summarized in Table~\ref{table:results}. The experimental results indicate more charge ejected and a higher conversion efficiency than we find in simulations.  We discuss possible reasons for this in Sec.~\ref{sec:discuss}. For the \relI/ case the discrepancy is less severe and the conversion efficiency from simulations is more similar to the lower bound on the conversion efficiency inferred from the experiment.
%%%%%%%%%%%%%%%%%%%%%%%%%%%%%%%%%%%%%%%%%%%%%%%%%%%%%%%%%%%%%%%%%%%%%%%%%%%%%%%%%%%%%%%%%
\section{Discussion}
\label{sec:discuss}
\subsection{Qualitative Differences Between Simulations}
All simulations presented here demonstrate significant numbers of super-ponderomotive electrons ejected backwards in the direction of the reflection of the normally incident pulse, where the ponderomotive potential for the linearly polarized simulations is 32~KeV for \mildI/ and 180~KeV for \relI/. This extends and confirms the results of earlier \twod/ simulations presented in \cite{Orban_etal2015} and it corroborates recent experimental measurements with an electron spectrometer indicating that significant numbers of MeV-scale electrons are ejected from the target \cite{Feister_etal2015}.

Table \ref{table:results} summarizes the results for the conversion efficiencies. It is interesting to compare the \twod/ and 3D simulation results. For the linear polarized simulations, the \twod/ and 3D simulations have conversion efficiencies that agree within 30\% for both simulated laser intensities. We had expected the 3D simulations to exhibit a significantly lower conversion efficiency from the simple consideration that 3D simulations include a third spatial dimension and electrons are no longer confined to the plane of the laser polarization. Thus we are currently without a satisfactory explanation for this result. We summarize the qualitative differences between \twod/ and 3D simulations in Sec.~\ref{sec:revisit}.

Although the conversion efficiencies are comparable, the angle-energy-spectra show interesting discrepancies between the \twod/ simulations and the analogous 3D simulations. For the \mildI/ simulations, note that the lower row, first and second columns of Fig.~\ref{fig:main_result} shows an increase in the angular spread, manifesting as a large amount of charge beyond the angular shadow of the OAP, while for the 3D simulations, most of the spectrum is focused in ejection angles directly backwards, almost precisely within the opening angle of the OAP. Meanwhile, for \relI/ results, the  \twod/ show more charges backwards ejected at an angle from 180$^\circ$ above 2 MeV, and almost no charge ejected directly backwards, whereas backwards is the significant ejection angle for charges ejected from the 3D simulations.

Fig.~\ref{fig:5e17_lpvcp} presents 3D simulation results for the circularly polarized and linearly polarized simulations which were performed with the \mildI/ intensity and the same spot size. There is a stark difference between the circularly polarized and linearly polarized results. For the circular polarized simulation, ejected electrons range from low energies until 3~MeV, while for the linearly polarized simulation, ejected electron energies only reach $\sim$1~MeV (Fig.~\ref{fig:5e17_lpvcp}). Interestingly, the electrons from the circular polarized simulation are preferentially ejected directly backwards from the target, rather than exhibiting preferred angles away from the back direction. This is likely due to the nature of circularly polarized laser pulses, which exhibit no preferred direction for the laser electric fields. By contrast, linearly polarized laser pulses do have a specified polarization angle and, as discussed in the next section and in Appendix~\ref{ap:analytics}, linearly polarized plane waves will preferentially direct electrons towards specific angles, depending on the electron energy. This does not occur for circularly polarized laser pulses and we are still working to incorporate circularly polarized light into the framework described in the next section. 

One possible explanation for this is that the reflectivity of circularly polarized light is be greater than that of linearly polarized light (both at normal incidence) which could result in back-directed electrons experiencing higher peak electric fields in the circularly polarized case.  However, we measured the reflectivity in the \mildI/ simulations and found that the reflectivity was similar in the circularly polarized and linearly polarized cases. The conversion efficiency and peak energy is very important for applications and we plan to continue to pursue a theoretical explanation for the enhanced electron acceleration in future work.

\subsection{An Analytic Model}
\label{sec:analytic}

To help understand the dynamics of electrons in simulation with linearly-polarized laser pulses and to interpret earlier results, we developed an analytic model for single particle motion that can provide an estimate for both the energy of the ejected electrons, and the direction of ejection. This model significantly improves upon \cite{Orban_etal2015} which only provided an order-of-magnitude estimate for the ejected electron energy.

The model is as follows: First, consider if an electron is ejected from the target in the region near the critical density by the standing-wave mechanism \cite{Kemp_etal2009,Orban_etal2015}, possibly with an extra boost from the electrostatic fields due to the evacuation of charges due to the ionization of the target by the incident pulse. For electrons ejected into the reflected pulse, we can consider, for simplicity, if the electrons then propagate according to the classical motion in a electromagnetic pulsed plane wave in vacuum, ignoring other effects such as quasi-static electric and magnetic fields, and the focusing (or defocusing) of the reflected laser pulse. The motion of a charged particle in a plane-wave in vacuum is a well-known result from Landau and Lifshitz \cite{LandauLifshitz}, in which electrons move with the quiver velocity in the polarization direction in response to the laser electric fields and in the forward direction in response to the laser magnetic fields and the Lorentz force. The well-known, parametric Landau \& Lifshitz result is often employed in the non-relativistic case when $a_0<1$. In the relativistic case in which the particle has an arbitrary initial velocity in any direction, and with an arbitrary time envelope, we derive the momenta and energies of charged particles in Appendix~\ref{ap:analytics} cast in a form that is parametric in the instantaneous phase of the laser as observed by the charged particle, $\eta=\omega t-\vec{k}\cdot\vec{x}$. We then choose a temporal pulse shape of a sine squared envelope, which matches the temporal envelope of our incident beam in our simulations. An important parameter in the model is the ratio of the pulse period (FWFM) to the laser period, which we denote as $\alpha = 60 \, \rm{fs} / 2.67 \, \rm{fs} \approx 22.5$.

It is well known that there is no net energy gain for an electron in a plane wave in vacuum\cite{Esarey_1995}. Moreover, it remains true that an electron overtaken by a time-pulsed plane wave in the absence of focusing will receive no net energy gain. %% Citation!!!!
However, if we assume that the injection of the electron into the reflected pulse is due to standing wave fields, a net energy gain will occur because the electron will be launched into the mid-way into the reflected pulse and not the beginning or ahead of the pulse.

%Should we include this additional statement.
%
%We refer to the instantaneous phase of the plane wave at the moment of injection at $\eta_0$. As described in the appendix, the final momentum gained by the electron is determined from the condition $\eta = 2 \pi \alpha$.

\begin{figure}
\includegraphics[angle=0,width=3.4in]{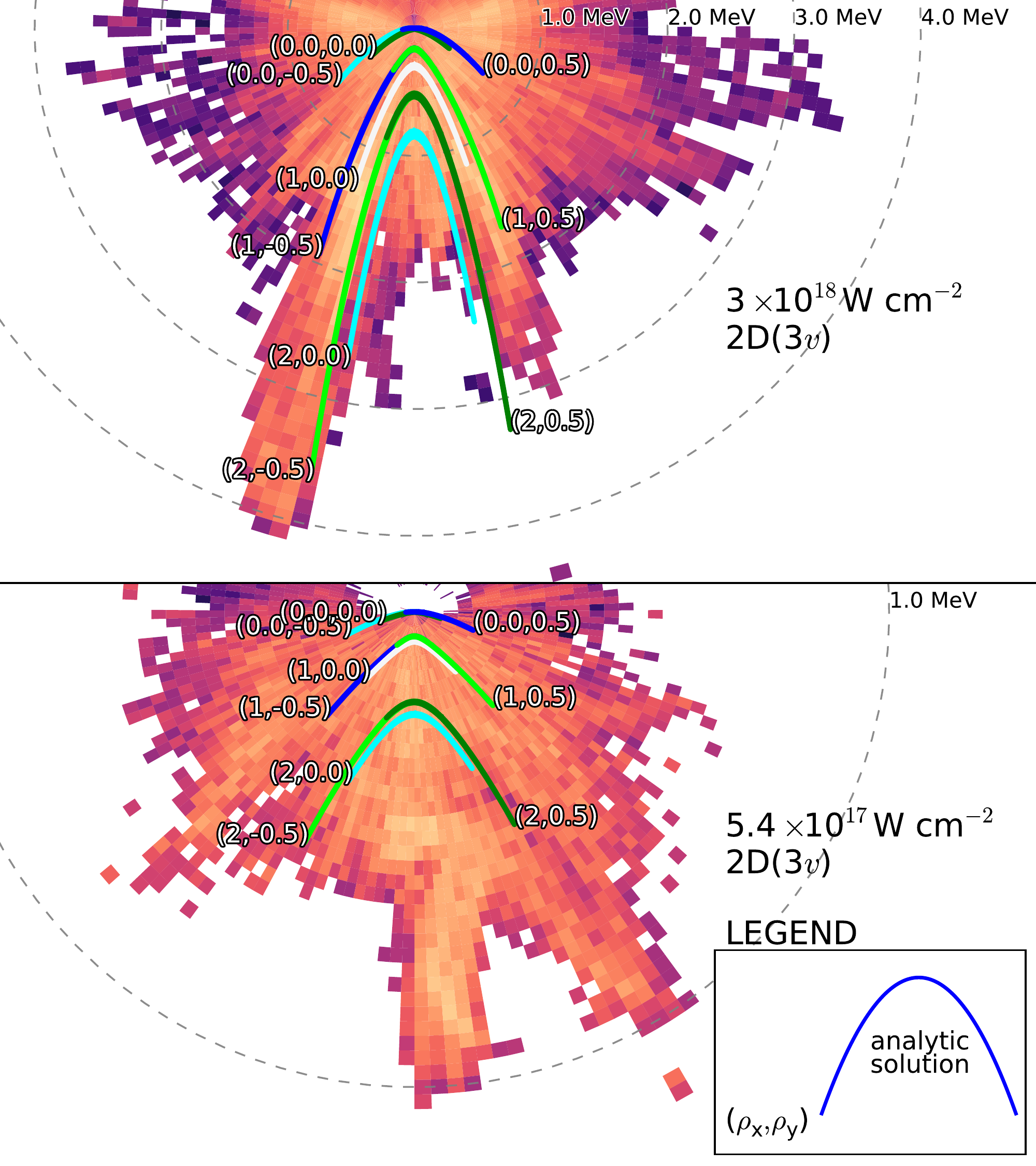}
\caption{Analytic, pulsed plane wave solutions from Appendix \ref{ap:analytics} overlaid with the \twod/ simulation angle-energy spectra, the first column of Fig.~\ref{fig:main_result}. The curved lines are plots of kinetic energy vs. ejection angle (here, in the polarization plane) with each color representing a particular injection momentum. Here the injection momentum is scaled to $a_0$. Specifically, each line corresponds to an injection momentum $\vec{p}/mc = \gamma_0\vec{\beta_0}~=~a_0 \vec{\rho}$ on injection into the reflected pulse. Each line is labelled by it's $\vec{\rho}=(\rho_x,\rho_y)$ value (as in the legend inset). The lines are created by varying the injection phase $\eta_0$ from $ \pi \alpha$ to 2$\pi \alpha$, corresponding to phase of the laser at the point and time of injection. For more specific details, refer to the Appendix.}
\label{fig:theory}
\end{figure}

The analytic solution from the Appendix yields expressions for the kinetic energy $(\gamma-1)mc^2$ where $\gamma$ is given by \ref{eq:gamma_soln} and the ejected angle $\phi$ in the polarization plane using $p_y/p_x=\tan\phi$, where the momentum from \ref{eq:p_soln} is utilized. From these two expressions, we can create parametric plots of energy vs. ejection angle that can be compared with earlier plots from the LSP simulations.  Fig.~\ref{fig:theory} shows this comparison for the \twod/ simulations with \mildI/ (top) and \relI/(bottom). In the model predictions in Fig.~\ref{fig:theory} we choose the incident momentum into the reflected pulse as up to $\sim a_0$ in the traverse directions and $\sim 2 a_0$ in the backwards (longitudinal) direction, as discussed in Appendix~\ref{ap:analytics}. Although it is a parametrized model, there appears to be reasonable choices that produce preferred angles away from the backwards direction that are similar to the preferred angles seen in the simulation results. The model also provides a more-precise explanation for how electrons are accelerated to super-ponderomotive energies than was presented in \cite{Orban_etal2015}.

\begin{figure}
\includegraphics[angle=0,width=3.2in]{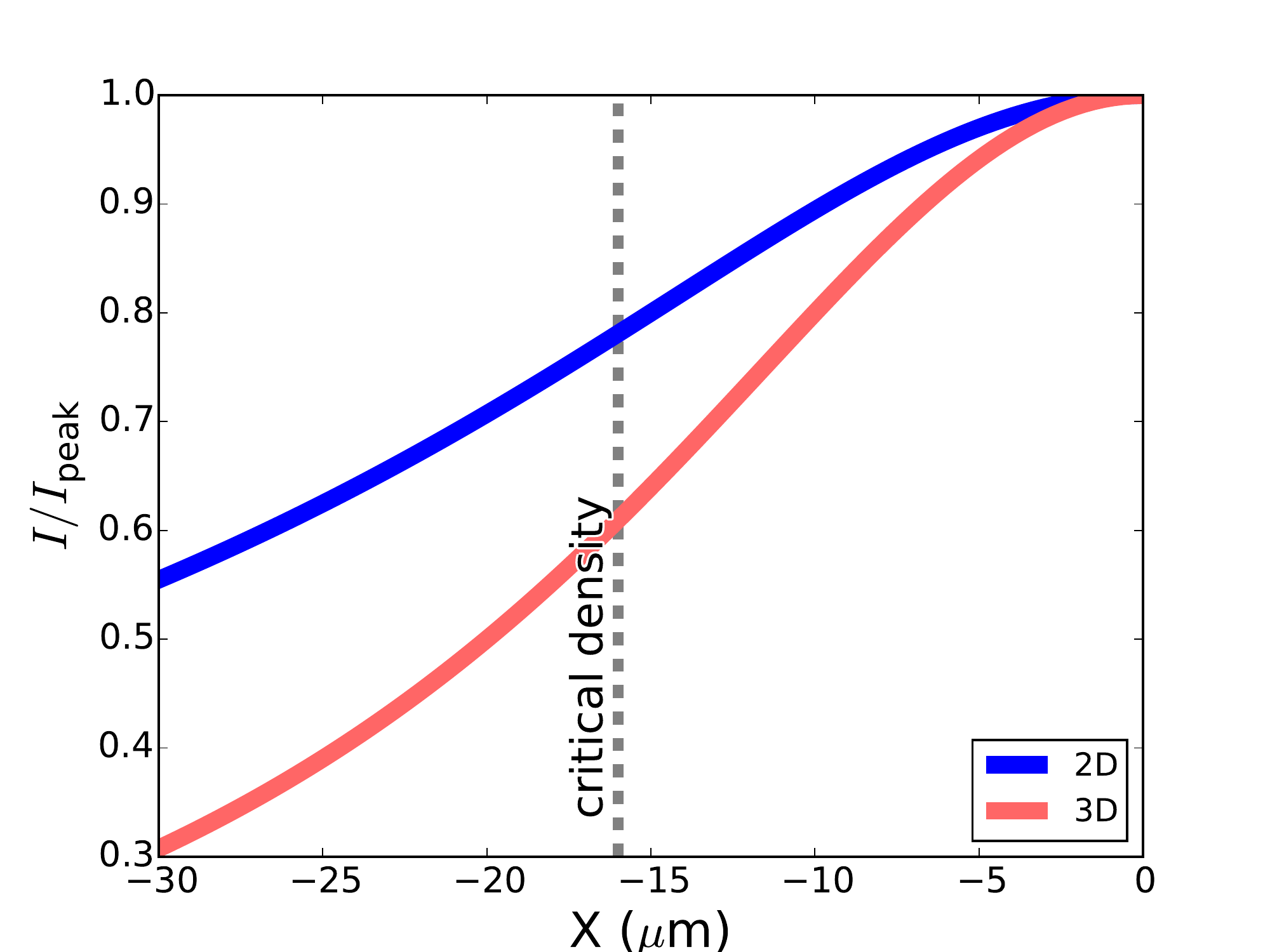}
\vspace{-0.2cm}
\caption{Identically-defined Gaussian lasers evolve to the same peak intensity in different ways between \twod/ and 3D Cartesian. In \twod/ PIC simulations, a Gaussian laser propagates as a 2D Gaussian ``wedge'' pulse while it propagates as a true Gaussian laser in 3D PIC simulations. This leads to intensity differences before the focus between \twod/ and 3D, even though both pulses reach the same peak intensity. The laser axis intensity of a Gaussian in 2D PIC~(blue) and 3D PIC~(red) in vacuum is plotted as a function propagation distance along $\vec{k}$ ($x$). The vertical dashed line marks the location of the critical density surface in our PIC simulations.}
\label{fig:3dv2d_focus}

\end{figure}

\subsection{\twod/ and 3D Results}
\label{sec:revisit}

Although the plane-wave model gives a possible explanation for some of the features seen in the angular-energy spectra, it does not explain the differences between the \twod/ and 3D simulations. Note that the form of Eq.~\ref{eq:p_soln},\ref{eq:gamma_soln} implies that for electrons in a plane wave, the maximum kinetic energy occurs when the transverse momentum is also maximized. Thus, one expects very few energetic electrons to be moving directly backwards ($\beta_y = 0$) according to this model. So while this model may provide some insight into the \twod/ simulations its utility may be limited in interpreting the 3D simulation results, which do not appear to exhibit preferred angles away from the backwards direction.

As mentioned earlier, the differences between \twod/ and 3D simulations were more pronounced for the \relI/ case. We do not yet have a satisfying explanation for this but it is worth outlining the known differences between \twod/ and 3D PIC simulations since there is likely multiple causes for the observed difference. Qualitative differences between 2D3($v$) and 3D simulations include:

\begin{enumerate}
  \item Electro-static fields decay as $1/r$ in \twod/ simulations instead of $1/r^2$ as would be the case in a truly 3D world.
  \item Particles are confined to the plane of the laser polarization in \twod/ instead of being free to move in all three spatial directions.
  \item Laser light comes to focus differently in \twod/ than would occur in 3D.
\end{enumerate}

Regarding the first point, \citet{Orban_etal2015} concluded that charge-separation effects caused by the overlap of the forward-going and reflected laser pulse will produce quasi-static electric fields that can enhance electron acceleration. These quasi-static electric fields develop in a process known as ``ponderomotive steepening''
\cite{Estabrook_Kruer1983}, which is so named because these fields will ultimately cause a steepening of the ion and electron density profiles. But before this occurs, quasi-static electric fields can provide an extra boost to some electrons that ultimately reach higher energies than they would without this effect \cite{Orban_etal2015}. Since the quasi-static electric fields decay with a different radial dependence in 3D and \twod/ simulations, this will lead to a different acceleration. With this in mind, in future work we plan to measure the precise energy gain from quasi-static electric fields in the simulations from analyzing particle trajectories.

The second point mentioned -- namely the confining of particles to the plane of the polarization -- affects the realism of the simulation in a number of ways. For example, the real laser pulse will produce gradients in the electron density both in and out of the plane of the polarization of the laser. There is also the possibility that a charge near the laser axis will experience larger electric and magnetic fields than it would otherwise because it is confined to the plane of polarization and cannot move out of the laser spot in the direction perpendicular to this plane. This latter consideration is another reason for surprise that the conversion efficiencies of the 3D simulations did not turn out to be significantly lower than the \twod/ simulations.

For an illustration of the third point, see Fig.~\ref{fig:3dv2d_focus}. This figure shows the spatial dependence of the laser intensity along the laser axis for the laser in vacuum as a function of the distance from peak focus ($x = 0$). This dependence is different in \twod/ simulations than it is in 3D because the \twod/ geometry can be thought of as a waveguide. This causes the light to come to focus less rapidly than a 3D laser pulse with the same spot size and intensity at peak focus. As a result the laser intensity near the critical density is substantially larger in the \twod/ simulations than in the 3D simulations. This could artificially enhance the conversion efficiency of \twod/ simulations relative to 3D, yet, as mentioned earlier, the conversion efficiencies of \twod/ and 3D simulations are overall quite similar.
Moreover, as we discuss in the Appendix \ref{ap:analytics}, the spatially escaping electrons could be considered as electrons escaping a foreshortened time pulse with a smaller value of $\alpha$, as long as they remain in the pulse for more two or three laser periods $(\alpha>3)$. For a tighter focus, more electrons will be free to escape the pulse, after which they will no longer propagate according to the pulsed wave model and other effects can become important. Regardless, this remains a puzzling result that we will return to in future work.

\section{Conclusions}

The results of \twod/ and 3D PIC simulations using the LSP code were compared in the context of electron acceleration using a high intensity, linearly-polarized laser pulse interacting with a water jet target. \citet{Orban_etal2015} presented the first \twod/ PIC simulations of this experiment (\citet{Morrison_etal2015} and \citet{Feister_etal2015}) and commented on the mechanisms of electron acceleration. In this paper we present 3D PIC simulations of this experiment for the first time and compare with \twod/ PIC simulations. These simulations were performed at intensities \mildI/ and \relI/ which bracket the experimental intensity ($\approx10^{18}$~W~cm$^{-2}$). 

Comparing \twod/ and 3D results, there are some differences in the angular distribution of ejected electrons, but both \twod/ and 3D simulations confirm significant back-directed electrons with super-ponderomotive energies, in agreement with the experiment. More quantitatively, while the laser-to-ejected-electron conversion efficiencies from \twod/ and 3D simulations with the same intensity and spot size were similar, the precise values were somewhat lower than the estimated lower bound conversion efficiency from experiment of 1.5\% for electrons with kinetic energies greater than 120~keV \cite{Morrison_etal2015}.

In an effort to understand the angular and energy distribution of ejected electrons we developed a parameterized analytic model that considers the dynamics of electrons that are accelerated initially by the standing wave fields and quasi-static electric fields present in the laser-interaction region. Later these electrons experience the reflected laser pulse which we model as a plane wave. We find that this model can describe a number of features observed in the angular-energy spectra of \twod/ simulations. 

We also performed a 3D PIC simulation using circular polarized laser light with the same parameters as the \mildI/ simulation. Remarkably, electron energies were observed up to $\sim$3~MeV and the conversion efficiency increased to 1.8\%. This result indicates that experiments with circularly polarized light should prove to be more effective than the experiments that have been conducted with linearly polarized light. The nature of this enhanced electron acceleration will be considered in future work.

%-----------------------------------------------------------------------------------------
%%%%%% UNCOMMENT FOR ARXIV
\section*{Acknowledgements}

%%%% COMMENT OUT FOR ARXIV
% \begin{acknowledgments}

%\acknowledgements

This research was sponsored by the Quantum and Non-Equilibrium Processes Division of the Air Force Office of Scientific Research, under the management of Dr. Enrique Parra, Program Manager. The authors acknowledge significant support from the Department of Defense High Performance Computing Modernization Program (DOD HPCMP) Internship Program. Supercomputer time was used on the DOD HPC Spirit and Garnet supercomputers. Resources were also used at the Ohio Supercomputer Center. 

%%%% COMMENT OUT FOR ARXIV
% \end{acknowledgments}

% %%%% BEGIN UNCOMMENT FOR ARXIV
\bibliography{ms.bib}
\bibliographystyle{apsrev}

% \onecolumngrid
% %%%% END UNCOMMENT FOR ARXIV

\appendix
\section{Energy and Motion For an Electron In a Pulsed Plane Wave}
\label{ap:analytics}

Consider an electron in a linearly polarized plane wave where $\vec{k}=k\hat{x}$ and $\hat{y}$ is the polarization direction, and in which the electric field has the form $\vec{E}(\vec{x},t) = E_0 f( \eta)\ \hat{y}$ where $\eta = \omega (t - x/c)$, $\omega$ is angular frequency of the plane wave in the lab frame, and $c$ is the speed of light. The Lorentz force can be written
\begin{equation}\label{eq:orig_lf}
    \frac{1}{\omega}\frac{d}{dt}(\gamma \vec{\beta}) = -a_0 f(\eta)~(\hat{y}(1-\beta_x) + \beta_y \hat{x})
\end{equation}
where we have expanded the $\vec{\beta} \times \hat{z}$ term and identified $a_0 = e E_0/ m \omega c$, where $-e$ is the charge of the electron, and $m$ is the electron's mass. $a_0$ is often referred to as the quiver velocity or the normalized vector potential. From the $\hat{x}$ component of ~\ref{eq:orig_lf} and the power equation one can obtain,
\begin{equation}\label{eq:const}
    \frac{d}{dt}(\gamma \beta_x) = -a_0\omega f(\eta)~\beta_y  = \frac{d}{dt}(\gamma \vec{\beta})\cdot \vec{\beta}=\frac{d}{dt}\gamma ,
\end{equation}
which implies that the doppler shift factor $\gamma - \gamma \beta_x = \gamma d \eta/d t$ is a conserved quantity.

Given that the force depends on the Lorentz invariant, but dynamic phase $\eta$, this motivates the change of variable from the frame time to the phase ($t \rightarrow \eta$). Noting that $d \eta /d t = \omega (1-\beta_x)$ and using chain rule yields the equation of motion
\begin{equation}\label{eq:trans_lf}
    \frac{d}{d \eta} (\gamma \vec{\beta}) = -a_0 f(\eta) \left( \hat{y} + \hat{x} \frac{\gamma\beta_y}{\gamma - \gamma \beta_x} \right).
\end{equation}
We can directly integrate the $\hat{y}$ component of~\ref{eq:trans_lf} and then, substitute the result of that integration into the $\hat{x}$ component, while noting that the denominator is constant due to \ref{eq:const}. To facilitate this derivation, we define the first structure function
\begin{equation}\label{eq:f1_def}
f_1(\eta_0,\eta) = \int_{\eta_0}^{\eta} d\eta' f(\eta') 
\end{equation}
and the second structure function
\begin{eqnarray*}\label{eq:f2_def}
f_2(\eta_0,\eta) &=& \int_{\eta_0}^{\eta} d\eta'f(\eta') f_1(\eta')\\
     &=& \int d f_1(\eta) f_1(\eta) \\
     &=& f_1(\eta_0,\eta_0)^2 / 2,
\end{eqnarray*}
where we simplified the second structure function and defined it in terms of the $f_1$. We thus obtain the following solution for $\gamma \vec{\beta}$ as a function the phase $\eta$:
\begin{multline}\label{eq:p_soln}
 \gamma\vec{\beta}(\eta) = \gamma_0\vec{\beta}_0 - \hat{y}~a_0 f_1(\eta_0,\eta)\\+\frac{\hat{x}}{\gamma-\gamma \beta_x} \bigg(\frac{a_0^2}{2}f_1(\eta_0,\eta)^2 - \gamma_0\beta_{y0}a_0f_1(\eta_0,\eta) \bigg).
\end{multline}
where $\gamma_0$ and $\vec{\beta}_0$ is the $\gamma$-factor and speed at an arbitrary starting phase  $\eta_0$. We will use $\eta_0$ as the phase at injection into the reflected pulse.
For the case where $f(\eta)=\cos \eta,\eta_0=0$, which corresponds to a simple plane wave, then $f_1(\eta_0=0,\eta)=\sin \eta$ which corresponds to the solution given in Landau and Lifshitz \cite{LandauLifshitz}.
 
Given that $\gamma_0 - \gamma_0\beta_{x0}=\gamma - \gamma\beta$, we obtain the energy for the electron
 \begin{equation}
 \gamma(\eta) = \gamma_0 + \frac{1}{\gamma - \gamma \beta_x}\frac{a_0^2}{2}\sin^2 \eta - \frac{\gamma_0\beta_{y0}}{\gamma - \gamma \beta_x} a_0 \sin\eta.
 \label{eq:gamma_soln}
 \end{equation}
Finally, note that 
\begin{eqnarray*}
    \textstyle \vec{\beta}  &= & \frac{1}{c}\frac{d\vec{x}}{dt}  = \frac{d k \vec{x}}{d\eta} (1-\beta_x) \\
    \Rightarrow k\vec{x} & = & \int_0^\eta d \eta~ \frac{\gamma \beta (\eta)}{\gamma -\gamma\beta_x},  \label{eq:speed}
\end{eqnarray*}
which, from \ref{eq:p_soln}, yields the (normalized) position as a function of $\eta$
\begin{multline}\label{eq:kx_soln}
 k \vec{x}(\eta) = k \vec{x}_0 + \frac{\gamma_0\vec{\beta}_0}{\gamma-\gamma \beta_x} \eta  - \hat{y}~\frac{a_0 f_{1i}(\eta_0,\eta)}{\gamma-\gamma \beta_x} \\
 + \frac{\hat{x}}{(\gamma - \gamma \beta_x)^2}
    \bigg(a_0^2 f_{2i}(\eta_0,\eta) - \gamma_0\beta_{y0} a_0f_{1i}(\eta_0,\eta) \bigg)
\end{multline}
where $f_{1i} = \int d\eta' f_1$ and $f_{2i} = \int d\eta' f_2 = \frac{1}{2}\int d\eta' f_1^2$. Finally, using that $\eta = \omega t - k x$, We obtain that $\omega t(\eta_0,\eta) = \eta + kx(\eta_0,\eta)$, where we use \ref{eq:kx_soln}. This gives all quantities as a function of frame time $\omega t(\eta)$ parametric in $\eta$.

Using \ref{eq:p_soln}, we define a pulsed plane wave that has a sine squared shape corresponding to the time envelope of our simulations. To do this, we set
\begin{equation}\label{eq:f_pulse}
f(\eta) = \sin \eta \sin \frac{\eta}{2\alpha} H(\eta) H(2\pi\alpha - \eta)
\end{equation}
where $H(x)$ is the Heaviside-step function where $H(x<0)=0$ and $H(x>0)=1$. Here, $\alpha$ is the ratio of the pulse time to the laser period. For our simulations, the pulse is 60 fs long (30 fs full-width at half-maximum), so $\alpha=22.48$. We omitted a pulse phase shift for simplicity. Using this choice for $f(\eta)$ yields the following structure functions and integrals:
\newcommand{\sinterms}[2][\sin]{#1 [\eta\frac{2\alpha #2 1}{2\alpha}]
-#1[\eta_0\frac{2\alpha #2 1}{2\alpha}]}

\begin{align}
\begin{split}
f_1(\eta_0, \eta) &= \frac{\sinterms{-}}{(2\alpha - 1)/\alpha}\label{eq:f1_pulse}\\
&\quad - \frac{\sinterms{+}}{(2\alpha + 1)/\alpha},
\end{split}\\
f_2(\eta_0, \eta) &= f_1(\eta_0,\eta)^2 / 2 \label{eq:f2_pulse},\\
\begin{split}
f_{1i}(\eta_0, \eta) &= -\frac{\sinterms[\cos]{-}}{(2\alpha -1)^2/2\alpha^2} \\
&\quad + \frac{\sinterms[\cos]{+}}{(2\alpha +1)^2/2\alpha^2}\\
&\quad - f_1(0,\eta_0)(\eta-\eta_0),\label{eq:f1i_pulse}
\end{split}
\end{align}
and, 
\newcommand{\sintwoterms}[1]{\frac{\alpha^2}{2 (2\alpha #1 1)^2}\bigg[\frac{\eta-\eta_0}{2} - \frac{\alpha}{2}\frac{\sin[\eta \frac{2\alpha #1 1}{\alpha}] - \sin[\eta_0 \frac{2\alpha #1 1}{\alpha}]}{2\alpha #1 1}\bigg]}
\begin{multline}\label{eq:f2i_pulse}
f_{2i}(\eta_0, \eta) = \\
\sintwoterms{-}\\
+\sintwoterms{+}\\
+\frac{\alpha^2}{4(4\alpha^2-1)} 
\bigg[ \sin 2 \eta - \sin 2 \eta_0 -
2 \alpha \bigg(\sin \frac{\eta}{\alpha} - \sin \frac{\eta_0}{\alpha}\bigg) \bigg] \\ 
+ f_1(0,\eta_0)[f_{1i}(\eta_0,\eta) - \frac{1}{2} f_1(0,\eta_0) (\eta-\eta_0)]
\end{multline}
For Eqs. \ref{eq:f1_pulse}-\ref{eq:f2i_pulse}, $\eta_0,\eta$ is restricted between $0$ and $2\pi \alpha$ due to the step functions in \ref{eq:f_pulse} that make the pulse finite. Given the mechanics of a standing wave, we expect that an electron injected from the standing wave into the reflected pulse will be injected for a given phase $\eta_0$ in the middle of the pulse and remain in the pulse until the pulse overtakes the electron. The end of the pulse as observed by the electron corresponds to the phase $\eta=2\pi\alpha$. This leads to a net energy gain for electrons in the reflected pulse. We note that for the case in which a pulse simply passes over an electron in free space, corresponding to $\eta_0=0,\eta=2\pi\alpha$, there is no net energy gain from \ref{eq:p_soln} except for moments when the electron is in the laser field \cite{Esarey_1995}.
Instead, superponderomotive energies are achieved because the standing wave injects electrons \emph{mid-way} into the reflected pulse and with a significant momentum. 
%the asymmetry due to the formation for standing wave, and the presence a plasma which results in the inducement of charge separation near the critical density surface \cite{Orban_etal2015} leads to injection into the middle of the reflected pulse and injection with a significant momentum. 

If the electron is injected at a phase before the middle of the pulse, i.e., $\eta_0<\pi \alpha$, then the symmetry of the pulse in time across the pulse's peak time will average out momentum gains across half the pulse. This warrants parametrizing $\eta_0$ from $\pi\alpha$ to $2\pi\alpha$. To create the theoretical model predictions shown in Fig.~\ref{fig:theory}, we take the ratio of $\gamma \beta_y/\gamma\beta_x=\tan \phi$ from \ref{eq:p_soln} with this range of $\eta_0$ and use \ref{eq:f1_pulse} for our model for $f_1$ to obtain the ejection angle of an accelerated electron in the polarization plane.  Likewise, we use these assumptions with \ref{eq:gamma_soln} to obtain the kinetic energy of an ejected electron. 

%We should note that some of the more extreme sets of initial conditions result in motion that test the limits of the applicability of this model, specifically we have found some initial conditions to predict motion well beyond the spot-size of the laser in which the spatial change of the intensity will result in ponderomotive force effects not included in this model. We should note though for an electron escaping the pulse not temporally but spatially, the fields observed by the electron will appear foreshortened, which at a first approximation, correspond to a shorter temporal pulsed plane wave within our model, corresponding to a smaller value for $\alpha$. We found insensitivity of the results to $\alpha$ granted the pulse encompassed more than three wavelengths, which seems to suggest that our model may still be applicable in the case of weak focusing, which we will revisit.

The model just described still requires an initial momentum, $\gamma_0 \vec{\beta_0}$, which we assume is on the order of the normalized electric field, $a_0$. It is well known that $a_0$ is the impulse done during one cycle by the transverse electric field of a simple plane wave. More importantly, $\sim a_0$ is an estimate for the longitudinal momentum from a standing wave as found by \cite{Kemp_etal2009}. According to \cite{Kemp_etal2009}, electrons that are ejected longitudinally from a plane wave can have a momentum significantly in excess of $a_0$, sometimes with as as much as $2 a_0$. For this reason, Fig.~\ref{fig:theory} shows this model's prediction for initial momenta scaled to $a_0$. Specifically, we consider initial momenta of the form $\vec{p}/m c = a_0 \vec{\rho}$, where we vary $\rho_x$ from $0$ to $2$ and $\rho_y$ between $-0.5$ and $0.5$.

We note that a weakness of the model is that for the intense case of \relI/ with large transverse injection momentum ($\rho_y\gtrsim0.5$), the model predicts transverse motion that would take the electron outside of the width of the laser pulse. This is not the case for the \mildI/ intensity. We show predictions that include $\rho_y = 0.5$ in Fig.~\ref{fig:theory} for both intensities for completeness even though the $\rho_y = \pm0.5$ cases with \relI/ are suspect. 
Conceivably, the energies and angles for $\rho_y = \pm 0.5$ and \relI/ may still be accurate for the simple reason that our results do not strongly depend on the precise value of $\alpha$. Electrons that travel outside of the laser width before the end of the pulse are similar to electrons that experience a pulsed plane wave with a smaller value of $\alpha$. Thus the model predictions for $\rho_y = \pm 0.5$ and \relI/ may still be qualitatively accurate.

%significantly beyond the width of the pulse. However, we suppose that for an electron escaping the pulse transversely, the fields observed by the electron will appear foreshortened, which at a first approximation, correspond to a shorter temporal pulsed plane wave within our model, corresponding to a smaller value for $\alpha$. We found insensitivity of the results to $\alpha$ granted the pulse encompassed more than three wavelengths, which seems to suggest that our model may still be applicable in the case of weak focusing of the \twod/ simulations.
%%% 

%%%% COMMENT OUT FOR ARXIV
% \bibliography{ms_PoP}

\end{document}